\documentclass[reprint,
superscriptaddress,
amsmath,amssymb,
aps,
prl,
]{revtex4-2}
\usepackage{graphicx}
\usepackage{dcolumn}
\usepackage{bm}
\usepackage{dsfont}
\usepackage{physics}

\begin{document}


\title{{Resolution of Gauge Ambiguities in Molecular Cavity Quantum Electrodynamics}}
\author{Michael A.D. Taylor}%
\affiliation{Department of Chemistry, University of Rochester, Rochester, NY 14627}
\affiliation{The Institute of Optics, Hajim School of Engineering, University of Rochester, Rochester, NY 14627}
\author{Arkajit Mandal}%
\affiliation{Department of Chemistry, University of Rochester, Rochester, NY 14627}
\author{Wanghuai Zhou}%
\affiliation{Department of Chemistry, University of Rochester, Rochester, NY 14627}
\affiliation{Advanced Functional Material and Photoelectric Technology Research Institution, School of Science, Hubei University of Automotive Technology, Shiyan, Hubei 442002, People's Republic of China}
\author{Pengfei Huo}
\email{pengfei.huo@rochester.edu}
\affiliation{Department of Chemistry, University of Rochester, Rochester, NY 14627}


\date{\today}
\begin{abstract}
This work provides the fundamental theoretical framework for the molecular cavity Quantum Electrodynamics by resolving the gauge ambiguities between the Coulomb gauge and the dipole gauge Hamiltonian under the electronic state truncation. Our conjecture for the arising of such gauge ambiguity is that not all operators are properly constrained in the truncated electronic subspace.  Based upon this conjecture, we construct a unitary transformation that properly constrains all operators in the subspace, and derive an equivalent and yet convenient expression for the Coulomb gauge Hamiltonian under the truncated subspace. We finally provide the analytical and numerical results of a model molecular system coupled to the cavity to demonstrate the validity of our theory.
\end{abstract}

\maketitle
Coupling molecules to the quantized radiation field inside an optical cavity creates a set of new photon-matter hybrid excitations, so-called polaritons~\cite{Flick2017PNAS,Ebbesen16,Feist2018,Ribeiro2018}. The rich dynamic interplay among these electronic, photonic, and nuclear degrees of freedom (DOF) has enabled a new paradigm for achieving unique chemical reactivities~\cite{Hutchison12,KowalewskiJCP2016,Galego2016,Mandal2019JPCL,Thomas2019}. The non-relativistic quantum electrodynamics (QED) Hamiltonian that describes such quantum light-matter interactions should obey the gauge principle, {\it i.e.}, giving rise to the same physical results (physical observables) upon a gauge transformation~\cite{Cohen-Tannoudji,Aspect}. While the QED Hamiltonian under both the Coulomb gauge and the dipole gauge (length gauge) indeed obeys this principle, these Hamiltonians under a finite electronic state truncation (the few-level approximation) are known to give different results for physical observable~\cite{Lamb1952,Aharonov,Lamb1987,Starace,Boyd,Rabl2018PRA2,Nori2019natphys,Nazir}, which is commonly referred to as the gauge ambiguity~\cite{Lamb1987,Boyd,Rabl2018PRA2,Nori2019natphys}. 
In this letter, we provide a conjecture for the cause of this ambiguity, and demonstrate that it is possible to resolve it, hence providing identical physical results from either the Coulomb or the dipole gauge Hamiltonian upon the same level of the electronic state truncation.

We begin by defining the matter Hamiltonian and the corresponding total dipole operator as follows
\begin{equation}\label{eqn:mu}
\hat{H}_\mathrm{M}=\hat{\bf T}+\hat{V}({\hat{\bf x}})=\sum_{j}\frac{1}{2m_j}\hat{\bf p}_{j}^2+\hat{V}({\hat{\bf x}});~\hat{\boldsymbol \mu}=\sum_{j}{z}_j \hat{\bf x}_j,
\end{equation}
where $j$ is the index of the $j_\mathrm{th}$ charged particle (including all electrons and nuclei), with the corresponding mass $m_j$ and charge ${z}_j$. In addition, $\hat{\bf x}\equiv\{\hat{\bf x}_{j}\}=\{\hat{\bf R},\hat{\bf r}\}$ with $\hat{\bf R}$ and $\hat{\bf r}$ represent the nuclear and electronic coordinates, respectively, $\hat{\bf p}\equiv\{\hat{\bf p}_{\bf R},\hat{\bf p}_{\bf r}\}\equiv\{\hat{\bf p}_{j}\}$ is the {\it mechanical} momentum operator as well as the canonical momentum operator, such that $\hat{\bf p}_j=-i\hbar{\boldsymbol\nabla}_{j}$. Further, $\hat{\bf T}=\hat{\bf T}_{\bf R}+\hat{\bf T}_{\bf r}$ is the kinetic energy operator, where $\hat{\bf T}_{\bf R}$ and $\hat{\bf T}_{\bf r}$ represent the kinetic energy operator for nuclei and for electrons, respectively, and $\hat{V}({\hat{\bf x}})$ is the potential operator that describes the Coulombic interactions among electrons and nuclei. 

The cavity photon field Hamiltonian under the single mode assumption is expressed as
\begin{equation}\label{phham}
\hat{H}_\mathrm{ph}=\hbar\omega_\mathrm{c}\big(\hat{a}^{\dagger}\hat{a}+\frac{1}{2}\big)=\frac{1}{2}\big(\hat{p}_\mathrm{c}^2+\omega_\mathrm{c}^2 \hat{q}_\mathrm{c}^2\big),
\end{equation}
where $\omega_\mathrm{c}$ is the frequency of the mode in the cavity,  $\hat{a}^{\dagger}$ and $\hat{a}$ are the photonic creation and annihilation operators, $\hat{q}_\mathrm{c} = \sqrt{\hbar/2\omega_\mathrm{c}}(\hat{a}^{\dagger} + \hat{a})$ and $\hat{p}_\mathrm{c} = i\sqrt{\hbar\omega_\mathrm{c}/2}(\hat{a}^{\dagger} - \hat{a})$ are the photonic coordinate and momentum operators, respectively. 
Choosing the Coulomb gauge, ${\boldsymbol \nabla} \cdot \hat{\bf A}=0$, the vector potential becomes purely transverse $\hat{\bf A} = \hat{{\bf A}}_\perp$. Under the long-wavelength approximation,
\begin{align}\label{eq:fields}
\hat{\bf A} = {\bf A}_0\big(\hat{a}+\hat{a}^{\dagger}\big)={\bf A}_0\sqrt{{2\omega_\mathrm{c}}/{\hbar}}~\hat{q}_\mathrm{c}
\end{align}
where ${\bf A}_0=\sqrt{\hbar/2 \omega_\mathrm{c} \varepsilon_{0} \mathcal{V}}~\hat{\bf e}$, with $\mathcal{V}$ as the quantization volume inside the cavity, $\varepsilon_0$ as the permittivity, and $\hat{\bf e}$ is the unit vector of the field polarization. 

The minimal coupling QED Hamiltonian in the {\it Coulomb} gauge (the ``$\mathrm{p\cdot A}$" form) is expressed as
\begin{equation}\label{eqn:Hc}
\hat{H}_\mathrm{C}=\sum_{j}\frac{1}{2m_j}(\hat{\bf p}_{j}-{z}_j\hat{\bf A})^2+\hat{V}(\hat{\bf x})+\hat{H}_\mathrm{ph},
\end{equation}
where $\hat{\bf p}_{j}=-i\hbar{\boldsymbol \nabla}_{j}$ is the {\it canonical} momentum operator. 
Upon a gauge transformation $\hat{U}_{\chi}=\exp[\frac{i}{\hbar}\sum_j z_j{\boldsymbol\chi}(\hat{\bf x}_j)]$, $\hat{H}_\mathrm{\chi}=\hat{U}_{\chi}\hat{H}_\mathrm{C}\hat{U}_{\chi}^{\dagger}$ remains gauge-invariant as
\begin{equation}\label{eqn:gauge-trans}
\hat{H}_\mathrm{\chi}=\sum_{j}\frac{1}{2m_j}(\hat{\bf p}_{j}-{z}_j\hat{\bf A}_{\chi}(\hat{\bf x}_j))^2+\hat{V}(\hat{\bf x})+\hat{H}^{\chi}_\mathrm{ph},
\end{equation}
where $\hat{H}^{\chi}_\mathrm{ph}=\hat{U}_{\chi} \hat{H}_\mathrm{ph}\hat{U}_{\chi}^{\dagger}$, $\hat{U}_{\chi}\hat{V}(\hat{\bf x})\hat{U}^{\dagger}_{\chi}=\hat{V}(\hat{\bf x})$ because $\hat{V}$ is a local potential operator for the matter, that is, only a function of $\hat{\bf x}$ and $\hat{\bf p}$-independent, and $\hat{\bf A}_{\chi}({\bf x}_j)=\hat{\bf A}+{\boldsymbol \nabla}_{j}\chi(\hat{\bf x}_j)$ is gauge transformed vector potential that provides the same physical field, because ${\boldsymbol\nabla}_{j}\times{\boldsymbol \nabla}_{j}\chi(\hat{\bf x}_j)=0$. To prove Eqn.~\ref{eqn:gauge-trans}, we have used $e^{\hat{Y}}\hat{O}(\hat{X})e^{-\hat{Y}}=\hat{O}(e^{\hat{Y}}\hat{X}e^{-\hat{Y}})$ for a unitary operator, as well as $\hat{U}_{\chi}\hat{\bf p}_{j}\hat{U}_{\chi}^{\dagger}=\hat{\bf p}_{j}-z_{j}{\boldsymbol\nabla}_{j}{\boldsymbol\chi(\hat{\bf x}_j)}$. 

We further introduce the Power-Zienau-Woolley (PZW) gauge transformation operator~\cite{PZW,GM-gauge,Cohen-Tannoudji} as
\begin{equation}\label{eqn:PZW}
\hat{U}=\exp \big[-\frac{i}{\hbar}\hat{\boldsymbol\mu}\cdot\hat{\bf A}\big]=\exp \big[-\frac{i}{\hbar}\hat{\boldsymbol \mu}\cdot{\bf A}_{0}\big(\hat{a}+\hat{a}^{\dagger}\big)\big],
\end{equation}
or $\hat{U}=\exp \big[-\frac{i}{\hbar}\sqrt{2\omega_\mathrm{c}/\hbar}\hat{\boldsymbol\mu}{\bf A}_0\hat{q}_\mathrm{c}\big]=\exp \big[-\frac{i}{\hbar} (\sum_{j}{z}_j \hat{\bf A}{\bf x}_j)\big]$.
Recall that a momentum boost operator $\hat{U}_\mathrm{p}= e^{-\frac{i}{\hbar} p_{0}\hat{q}}$ displaces $\hat{p}$ by the amount of $p_0$, such that $\hat{U}_\mathrm{p} \hat{O}(\hat{p}) \hat{U}_\mathrm{p}^\dagger = \hat{O}(\hat{p} + p_0)$. Hence, $\hat{U}$ is a boost operator for both the photonic momentum $\hat{p}_\mathrm{c}$ by the amount of $\sqrt{2\omega_\mathrm{c}/\hbar}\hat{\boldsymbol \mu}{\bf A_0}$, as well as for the matter momentum $\hat{\bf p}_{j}$ by the amount of ${z}_j \hat{\bf A}$. 
The PZW gauge operator (Eqn.~\ref{eqn:PZW}) is a special case of $\hat{U}_{\chi}$, such that ${\boldsymbol\chi}=-\hat{\bf x}_j\cdot\hat{\bf A}$.
Using $\hat{U}^{\dagger}$ to boost the matter momentum, one can show that
\begin{equation}\label{boosthm}
\hat{H}_\mathrm{C}=\hat{U}^{\dagger}\hat{H}_\mathrm{M}\hat{U}+\hat{H}_\mathrm{ph},
\end{equation}
hence $\hat{H}_\mathrm{C}$ can be obtained~\cite{Nori2019natphys} by a momentum boost with the amount of $-{z}_j \hat{\bf A}$ for $\hat{\bf p}_{j}$, then adding $\hat{H}_\mathrm{ph}$. 

The QED Hamiltonian under the {\it dipole} gauge (the ``$\mathrm{d\cdot E}$" form~\cite{PZW,GM-gauge}) can be obtained by performing the PZW transformation on $\hat{H}_\mathrm{C}$ as follows 
\begin{align}\label{eqn:ddote}
\hat{H}_\mathrm{D}&=\hat{U}\hat{H}_\mathrm{C}\hat{U}^\dagger=\hat{U} \hat{U}^{\dagger}\hat{H}_\mathrm{M}\hat{U}\hat{U}^{\dagger} +\hat{U}\hat{H}_\mathrm{ph}\hat{U}^\dagger\\
&=\hat{H}_\mathrm{M}+ \hbar \omega_\mathrm{c} (\hat{a}^\dagger \hat{a} + \frac{1}{2}) + i\omega_\mathrm{c} \hat{\boldsymbol \mu} {\bf A}_0 (\hat{a}^\dagger - \hat{a}) + \frac{\omega_\mathrm{c}}{\hbar}(\hat{\boldsymbol \mu}{\bf A}_0)^2 \nonumber,
\end{align}
where we have used Eqn.~\ref{boosthm} to express $\hat{H}_\mathrm{C}$, and the last three terms of the above equation are the results of $\hat{U}\hat{H}_\mathrm{ph}\hat{U}^\dagger$ (see SI). The above result is also apparent by examining Eqn.~\ref{eqn:gauge-trans}, where the gauge DOF $-z_j{\boldsymbol\nabla} {\boldsymbol\chi}(\hat{\bf x}_j)=z_j\hat{\bf A}$ exactly cancels the vector potential, leaving no explicit gauge-dependent quantity in $\hat{H}_\mathrm{D}$. Using $\hat{q}_\mathrm{c}$ and $\hat{p}_\mathrm{c}$, one can instead show that $\hat{H}_\mathrm{D}=\hat{H}_\mathrm{M}+\frac{1}{2}\omega_\mathrm{c}^{2}\hat{q}_\mathrm{c}^{2}+\frac{1}{2}(\hat{p}_\mathrm{c}+\sqrt{2\omega_\mathrm{c}/\hbar}\hat{\boldsymbol \mu}{\bf A}_0)^2$, because the PZW operator boosts the photonic momentum $\hat{p}_\mathrm{c}$ by $\sqrt{2\omega_\mathrm{c}/\hbar}\hat{\boldsymbol \mu}{\bf A}_0$. 

The gauge invariance is explicitly enforced between $\hat{H}_\mathrm{C}$ (Eqn.~\ref{eqn:Hc}) and $\hat{H}_\mathrm{D}$ (Eqn.~\ref{eqn:ddote}) through the unitary PZW Gauge transformation (Eqn.~\ref{eqn:PZW}). However, this gauge invariance will explicitly breakdown when a truncation of electronic states is applied to both Hamiltonians~\cite{Rabl2018PRA2,Nori2019natphys}. Consider a finite subset of electronic states $\{|\alpha\rangle\}$, where the projection operator $\hat{\mathcal P}=\sum_{\alpha}|\alpha\rangle \langle \alpha|$ defines the truncation of the full electronic Hilbert space $\hat{\mathds{1}}_\mathrm{r}=\hat{\mathcal P}+\hat{\mathcal Q}$ to the corresponding subspace $\hat{\mathcal P}$. 
The truncation reduces the size of the Hilbert space from originally $\hat{\mathds{1}}_\mathrm{r}\otimes \hat{\mathds{1}}_\mathrm{R}\otimes \hat{\mathds{1}}_\mathrm{ph}$ to now $\hat{\mathcal P}\otimes \hat{\mathds{1}}_\mathrm{R}\otimes \hat{\mathds{1}}_\mathrm{ph}$, where $\hat{\mathds{1}}_\mathrm{R}$ and $\hat{\mathds{1}}_\mathrm{ph}$ represent the identity operator of the nuclear and the photonic DOF, respectively. 
The truncated matter Hamiltonian is
\begin{equation}
\hat{\mathcal H}_\mathrm{M}=\hat{\mathcal P}\hat{H}_\mathrm{M}\hat{\mathcal P}=\hat{\mathcal P}\hat{\bf T}\hat{\mathcal P}+\hat{\mathcal P}\hat{V}(\hat{\bf x})\hat{\mathcal P}.
\end{equation}
Throughout this letter, we use calligraphic symbols (such as $\hat{\mathcal H}_\mathrm{M}$) to indicate operators in the truncated Hilbert space. Truncating the momentum operator and dipole operator as $\hat{\mathcal P}\hat{\bf p}_j\hat{\mathcal P}$ and $\hat{\mathcal P}\hat{\boldsymbol \mu}\hat{\mathcal P}$, the QED Hamiltonians under the truncated subspace are commonly defined as
\begin{align}
\hat{\mathcal H}'_\mathrm{C}&= \hat{\mathcal P}\hat{H}_\mathrm{C}\hat{\mathcal P}=\hat{\mathcal P}\hat{U}^{\dagger}\hat{H}_\mathrm{M}\hat{U}\hat{\mathcal P}+\hat{H}_\mathrm{ph} \label{eqn:hc'1} \\ 
&=\hat{\mathcal H}_\mathrm{M}+\hat{H}_\mathrm{ph}+\sum_{j}\big(-\frac{{z}_j}{m_j}\hat{\mathcal P}\hat{\bf p}_{j}\hat{\mathcal P}\hat{\boldsymbol A}+\frac{{z}_j^2\hat{\boldsymbol A}^2}{2 m_j}\big) \nonumber \\
\hat{\mathcal H}_\mathrm{D}&=\hat{\mathcal H}_\mathrm{M}+ \hat{H}_\mathrm{ph} + i \omega_\mathrm{c} \hat{\mathcal P}\hat{\boldsymbol \mu}\hat{\mathcal P} {\bf A}_0 (\hat{a}^\dagger - \hat{a}) + \frac{\omega_\mathrm{c}}{\hbar}(\hat{\mathcal P}\hat{\boldsymbol\mu}\hat{\mathcal P}{\bf A}_0)^2.\label{eqn:hd1}
\end{align}
It is a well-known fact that the above two Hamiltonians do not generate the same polariton eigenspectrum~\cite{Rabl2018PRA2,Nori2019natphys,Nazir} under the ultra-strong coupling regime~\cite{Kockum2019}, hence explicitly breakdown the gauge invariance, leading to the ambiguities~\cite{Lamb1987,Boyd} of using which Hamiltonian, $\hat{\mathcal H}'_\mathrm{C}$ or $\hat{\mathcal H}_\mathrm{D}$, to compute the physical quantities when applying to a state truncation through $\hat{\mathcal P}$. This is attributed~\cite{Rabl2018PRA2,Li2020} to the fact that $\hat{\mathcal H}'_\mathrm{C}$ usually requires a larger subset of the matter states to converge or generate consistent results of $\hat{\mathcal H}_\mathrm{D}$ and apparently, under the {\it complete} basis limit, they should be gauge invariant. Further, this fundamentally different behavior of $\hat{\mathcal H}'_\mathrm{C}$ and $\hat{\mathcal H}_\mathrm{D}$ upon electronic states truncation is also attributed to the fundamental asymmetry of the $\hat{\bf p}$ and $\hat{\boldsymbol\mu}=\sum_{j}\hat{\bf x}_{j}$ operators, especially when the potential $\hat{V}(\hat{\bf x})$ is highly anharmonic~\cite{Rabl2018PRA2}.

We conjecture this gauge-ambiguity is caused by the fact that $\hat{\mathcal P}\hat{U}^{\dagger}$ and $\hat{U}\hat{\mathcal P}$ operators used to construct $\hat{\mathcal H}'_\mathrm{C}$ (Eqn.~\ref{eqn:hc'1}) do not properly constrain all operators in the subspace $\hat{\mathcal P}$, such that some of them are entering into the subspace $\hat{\mathcal Q}=\hat{\mathds{1}}_\mathrm{r}-\hat{\mathcal P}$. Indeed, this is the case for $\hat{U}\hat{\mathcal P}=(\hat{\mathcal P}+\hat{\mathcal Q})\hat{U}\hat{\mathcal P}$. This is almost apparent by examining the diamagnetic term ${z}_j^2\hat{\boldsymbol A}^2/2 m_j$ in $\hat{\mathcal H}'_\mathrm{C}$, which is effectively evaluated in the full space $\hat{\mathds{1}}_\mathrm{r}$ (based on the Thomas-Reiche-Kuhn sum rule), hence is not properly confined in $\hat{\mathcal P}$. This diamagnetic term overestimates what it should be in the subspace~\cite{Rabl2018PRA2,Nori2019natphys}, and by confining it inside $\hat{\mathcal P}$, the results can be significantly improved~\cite{Nori2019natphys}. Note that using $\hat{\mathcal P}\hat{U}\hat{\mathcal P}=\sum_{n=1}^{\infty}\frac{1}{n!}(-\frac{i}{\hbar})^n \hat{\mathcal P}\hat{\boldsymbol\mu}^n \hat{\mathcal P} \hat{\bf A}^n$ does not resolve this problem either (as oppose to the original claim in Ref.~\citenum{Nori2019natphys}), because for all non-linear terms in $\hat{\mathcal P}\hat{U}\hat{\mathcal P}$ (for $n\ge 2$), $\hat{\mathcal P}\hat{\boldsymbol\mu}^n \hat{\mathcal P}$ is not properly confined in the $\hat{\mathcal P}$ subspace~\cite{Nazir}. For example, $\hat{\mathcal P}\hat{\boldsymbol\mu}^2\hat{\mathcal P}=\hat{\mathcal P}\hat{\boldsymbol\mu} (\hat{\mathcal P}+\hat{\mathcal Q})\hat{\boldsymbol\mu}\hat{\mathcal P}=(\hat{\mathcal P}\hat{\boldsymbol\mu}\hat{\mathcal P})^2 + \hat{\mathcal P}\hat{\boldsymbol\mu}\hat{\mathcal Q}\hat{\boldsymbol\mu}\hat{\mathcal P} \neq (\hat{\mathcal P}\hat{\boldsymbol\mu}\hat{\mathcal P})^2$, where we have used $\hat{\mathcal P}\hat{\mathcal P}=\hat{\mathcal P}$. Nevertheless, using $\hat{\mathcal P}\hat{U}\hat{\mathcal P}$ still provide slightly better numerical results~\cite{Nazir} compared to using $\hat{U}\hat{\mathcal P}$, because the former has a fewer terms in $\hat{\mathcal Q}$, corroborating our conjecture. Further, this conjecture also provides the reason why $\hat{\mathcal H}_\mathrm{D}$ needs to be constructed in the form of Eqn.~\ref{eqn:hd1}, because by doing so, all of the operators are properly confined in the $\hat{\mathcal{P}}$ subspace. Indeed, $\hat{\mathcal H}_\mathrm{D}$ gives very accurate results compared to the exact solution when a few level truncation is a valid approximation~\cite{Rabl2018PRA2}. Note that $\hat{\mathcal H}_\mathrm{D}\neq\hat{\mathcal P}\hat{H}_\mathrm{D}\hat{\mathcal P}$, because $\hat{\mathcal P}\hat{H}_\mathrm{D}\hat{\mathcal P}$ also has an operator (the Dipole self-energy term) $\hat{\mathcal P}\hat{\boldsymbol\mu}^2\hat{\mathcal P}$ that is not properly confined inside $\hat{\mathcal P}$, and will indeed give much worse numerical results than $\hat{\mathcal H}_\mathrm{D}$. Numerical examples are provided in SI to support our conjecture.

Based on the above conjecture, the gauge ambiguity will be resolved by defining the following unitary operator
\begin{equation}\label{eqn:epzw}
\hat{\mathcal U}=\exp \big[-\frac{i}{\hbar}\hat{\mathcal P}\hat{\boldsymbol\mu}\hat{\mathcal P}\cdot\hat{\bf A}\big]\equiv\exp \big[-\frac{i}{\hbar}\tilde{\boldsymbol\mu}(\hat{\bf x},\hat{\bf p})\cdot\hat{\bf A}\big],
\end{equation}
such that all terms in $\hat{\mathcal U}=\sum_{n=0}^{\infty}\frac{1}{n!}(-\frac{i}{\hbar})^n (\hat{\mathcal P}\hat{\boldsymbol\mu} \hat{\mathcal P})^n \hat{\bf A}^n$ are properly confined within the subspace $\hat{\mathcal P}$. Note that $\hat{\mathcal U}$ is defined in analogous to the PZW gauge operator (Eqn.~\ref{eqn:PZW}) in the full space, and $\hat{\mathcal P}\hat{\boldsymbol \mu}\hat{\mathcal P}\equiv\tilde{\boldsymbol\mu}(\hat{\bf x},\hat{\bf p})$ in principle is a function of both $\hat{\bf x}$ and $\hat{\bf p}$, due to the finite level projection that ruins the locality of $\hat{\bf x}$~\cite{Nori2019natphys,Nori2020}. Further, $\hat{\mathcal U}$ is a unitary transformation operator in the $\hat{\mathcal P}$ subspace and the identity operator in the subspace of $\hat{\mathds{1}}_\mathrm{r}-\hat{\mathcal P}$, such that we still have $\hat{\mathcal U}\hat{\mathcal U}^{\dagger}=\hat{\mathds{1}}_\mathrm{r}\otimes \hat{\mathds{1}}_\mathrm{R}\otimes \hat{\mathds{1}}_\mathrm{ph}=\hat{U}\hat{U}^{\dagger}$. 

Using the above operator $\hat{\mathcal U}$, one can define the following Coulomb gauge Hamiltonian
\begin{equation}\label{hc}
\hat{\mathcal{H}}_\mathrm{C}=\hat{\mathcal U}^{\dagger}\hat{\mathcal H}_\mathrm{M}\hat{\mathcal U}+\hat{H}_\mathrm{ph},
\end{equation}
in analogous to $\hat{H}_\mathrm{C}$ in Eqn.~\ref{boosthm} in the full space. One can then formally show that $\hat{\mathcal{H}}_\mathrm{C}$ (Eqn.~\ref{hc}) and $\hat{\mathcal{H}}_\mathrm{D}$ (Eqn.~\ref{eqn:hd1}) are related through $\hat{\mathcal U}$ (Eqn.~\ref{eqn:epzw}) as follows
\begin{equation}\label{eqn:hd2}
\hat{\mathcal U}\hat{\mathcal{H}}_\mathrm{C}\hat{\mathcal U}^{\dagger}=\hat{\mathcal H}_\mathrm{M}+\hat{\mathcal U}\hat{H}_\mathrm{ph}\hat{\mathcal U}^{\dagger}=\hat{\mathcal H}_\mathrm{D}, 
\end{equation}
as well as $\hat{\mathcal U}^{\dagger}\hat{\mathcal H}_\mathrm{D}\hat{\mathcal U}=\hat{\mathcal H}_\mathrm{C}$. Note that to establish the last equality in Eqn.~\ref{eqn:hd2} we have used the fact that $\hat{\mathcal U}\hat{H}_\mathrm{ph}\hat{\mathcal U}^{\dagger}=\hat{\mathcal U}(\frac{1}{2}\omega_\mathrm{c}^{2}\hat{q}_\mathrm{c}^{2}+\frac{1}{2}\hat{p}_\mathrm{c}^{2})\hat{\mathcal U}^{\dagger}=\frac{1}{2}\omega_\mathrm{c}^{2}\hat{q}_\mathrm{c}^{2}+\frac{1}{2}(\hat{p}_\mathrm{c}+\sqrt{2\omega_\mathrm{c}/\hbar}\hat{\mathcal P}\hat{\boldsymbol\mu}\hat{\mathcal P}{\bf A}_0)^2$. Thus, we have formally demonstrated that the gauge ambiguities between the Coulomb and dipole gauge Hamiltonians can be resolved for an arbitrary matter-cavity hybrid system, under the same level of electronic state truncation.

To derive the detailed expression of $\hat{\mathcal{H}}_\mathrm{C}$, we notice that $\hat{\mathcal P}\hat{\boldsymbol \mu}\hat{\mathcal P}\equiv\tilde{\boldsymbol\mu}(\hat{\bf x},\hat{\bf p})$ is in principle a non-linear function of both $\hat{\bf x}$ and $\hat{\bf p}$, as oppose to $\hat{\boldsymbol \mu}$ (see Eqn.~\ref{eqn:mu}) which is a pure linear function of $\hat{\bf x}$ . Thus, $\hat{\mathcal U}^{\dagger}$ no longer just boosts the matter momentum by ${z}_j \hat{\bf A}$. Using the Baker-Campbell-Hausdorff (BCH) identity,
we have
\begin{align}\label{eqn:boostp}
\hat{\mathcal U}^{\dagger}\hat{\bf p}_{j}\hat{\mathcal U}&=\hat{\bf p}_j+\frac{i}{\hbar}[\tilde{\boldsymbol\mu}\hat{\bf A},\hat{\bf p}_{j}]+\frac{1}{2}\big(\frac{i}{\hbar}\big)^2[\tilde{\boldsymbol\mu}\hat{\bf A},[\tilde{\boldsymbol\mu}\hat{\bf A},\hat{\bf p}_j]]+...\nonumber \\
&=\hat{\bf p}_{j}-{\boldsymbol\nabla}_{j}{\tilde{\boldsymbol\mu}(\hat{\bf x},\hat{\bf p})}\hat{\bf A}+{\bf \tilde P}_{j},
\end{align}
where ${\bf \tilde P}_{j}\equiv\frac{1}{2}\big(\frac{i}{\hbar}\big)^2[\tilde{\boldsymbol\mu}\hat{\bf A},[\tilde{\boldsymbol\mu}\hat{\bf A},\hat{\bf p}_j]]+...$ is the residual momentum that accounts for terms with more than one commutator in the BCH identity. Hence, under the projection, $\hat{\mathcal U}^{\dagger}$ boosts the matter momentum by the amount of $-{\boldsymbol\nabla}_{j}\tilde{\boldsymbol\mu}\hat{\bf A}+{\bf \tilde P}_{j}$. Similarly, $\hat{\mathcal P}\hat{V}(\hat{\bf x})\hat{\mathcal P}$ is in principle a non-local potential~\cite{Starace,Nori2019natphys,Nori2020}, such that $\hat{\mathcal P}\hat{V}(\hat{\bf x})\hat{\mathcal P}=\hat{\mathcal V}(\hat{\bf x},\hat{\bf p})$. Hence, $\hat{\mathcal U}^{\dagger}$ also displaces the matter coordinate as well as boost the matter momentum inside $\hat{\mathcal V}(\hat{\bf x},\hat{\bf p})$, whereas $\hat{\mathcal U}^{\dagger}\hat{\mathcal V}\hat{\mathcal U}$ can be formally derived through the BCH identity.

Using Eqn.~\ref{eqn:boostp}, as well as the fact that both $\hat{\mathcal U}^{\dagger}$ and $\hat{\mathcal U}$ commute with $\hat{\mathcal P}$, such that $\hat{\mathcal U}^{\dagger}\hat{\mathcal P}\hat{\bf T}\hat{\mathcal P}\hat{\mathcal U}=\hat{\mathcal P}\hat{\mathcal U}^{\dagger}\hat{\bf T}\hat{\mathcal U}\hat{\mathcal P}$, we can derive $\hat{\mathcal{H}}_\mathrm{C}$ (in Eqn.~\ref{hc}) as follows 
\begin{align}\label{hc5}
&\hat{\mathcal{H}}_\mathrm{C}=\hat{\mathcal U}^{\dagger}\hat{\mathcal P}\hat{\bf T}\hat{\mathcal P}\hat{\mathcal U}+\hat{\mathcal U}^{\dagger}\hat{\mathcal P}\hat{V}(\hat{\bf x})\hat{\mathcal P}\hat{\mathcal U}+\hat{H}_\mathrm{ph}\\
&=\sum_{j}\frac{1}{2m_j}\hat{\mathcal P}\big(\hat{\bf p}_{j}-{\boldsymbol\nabla}_{j} \tilde{\boldsymbol\mu}\hat{\bf A}+{\bf \tilde P}_{j})^2\hat{\mathcal P}+\hat{\mathcal U}^{\dagger}\hat{\mathcal V}(\hat{\bf x},\hat{\bf p})\hat{\mathcal U}+\hat{H}_\mathrm{ph},\nonumber
\end{align}
where the sum $j$ includes all charged particles (electrons and nuclei). Note that $\hat{\mathcal H}'_\mathrm{C}$ (Eqn.~\ref{eqn:hc'1}) as well as $\hat{H}_\mathrm{C}$ (Eqn.~\ref{eqn:Hc}) only contain the vector potential $\hat{\bf A}$ up to the second order. This is no longer the case for $\hat{\mathcal H}_\mathrm{C}$ in Eqn.~\ref{hc5}. In fact, both the ${\bf \tilde P}_{j}$ term and the $\hat{\mathcal U}^{\dagger}\hat{\mathcal V}(\hat{\bf x},\hat{\bf p})\hat{\mathcal U}$ term in principle contain infinite orders of $\hat{\bf A}$. 
It is also self-evident that $\hat{\mathcal{H}}_\mathrm{C}$ (Eqn.~\ref{hc5}) will return to $\hat{H}_\mathrm{C}$ (Eqn.~\ref{eqn:Hc}) under the complete basis limit, such that $\tilde{\boldsymbol\mu}\equiv\hat{\mathcal P}\hat{\boldsymbol\mu}\hat{\mathcal P}\to \hat{\boldsymbol\mu}$, thus ${\boldsymbol\nabla}_{j}\tilde{\boldsymbol\mu}\to {\boldsymbol\nabla}_{j}\hat{\boldsymbol\mu}={z}_j$, hence ${\bf \tilde P}_j\to 0$, as well as $\hat{\mathcal U}\to\hat{U}$, hence $\hat{\mathcal U}^{\dagger}\hat{\mathcal P}\hat{V}(\hat{\bf x})\hat{\mathcal P}\hat{\mathcal U}\to\hat{U}^{\dagger}\hat{V}(\hat{\bf x})\hat{U}=\hat{V}(\hat{\bf x})$. Unfortunately, $\hat{\mathcal{H}}_\mathrm{C}$ no longer remains a gauge-invariant form as indicated by Eqn.~\ref{eqn:gauge-trans}, except when approaching to the complete basis limit, and of course, $\hat{\mathcal H}_\mathrm{C}$ is invariant from $\hat{\mathcal H}_\mathrm{D}$ through the $\hat{\mathcal U}$ transformation as indicated in Eqn.~\ref{eqn:hd2}. 
Thus, $\hat{\mathcal H}_\mathrm{C}$ resolves the discrepancy (ambiguity) between the coulomb and dipole gauge by under matter state truncation, with the price that $\hat{\mathcal H}_\mathrm{C}$ is no longer gauge invariant in general.

To further present an equivalent yet convenient $\hat{\mathcal{H}}_\mathrm{C}$ for molecular cavity QED, we use the electronic states associated with the electronic Hamiltonian $\hat{H}_\mathrm{el}=\hat{\bf T}_{\bf r}+\hat{V}=\hat{H}_\mathrm{M}-\hat{\bf T}_{\bf R}$. The {\it adiabatic} electronic states $|\alpha ({\bf R})\rangle$ are one type of the most commonly obtained electronic states by solving $\hat{H}_\mathrm{el}|\alpha({\bf R})\rangle=(\hat{\bf T}_{\bf r}+\hat{V})|\alpha ({\bf R})\rangle=E_{\alpha} ({\bf R}) |\alpha ({\bf R})\rangle$, where $E_{\alpha} ({\bf R})$ is the so-called the adiabatic potential energy surface. Using a finite set of $\{|\alpha({\bf R})\rangle \}$ to form $\hat{\mathcal P}=\sum_{\alpha}|\alpha ({\bf R})\rangle\langle \alpha({\bf R})|$, the projected electronic Hamiltonian is $\hat{\mathcal H}_\mathrm{el}=\hat{\mathcal P}\hat{H}_\mathrm{el}\hat{\mathcal P}=\sum_{\alpha}E_{\alpha}({\bf R}) |\alpha\rangle\langle \alpha|$. Alternatively, diabatic electronic states~\cite{Mead:1982,Cederbaum:1989,cave1996,Yarkony:2016} $\{|\varphi\rangle,|\phi\rangle\}$ can be obtained by the unitary transform~\cite{Mead:1982,Cederbaum:1989,cave1996,Yarkony:2016,SubotnikDIA} from the adiabatic states $|\alpha ({\bf R})\rangle$ or by constructions through preserving basic properties~\cite{DiabaticARPC}. As oppose to adiabatic states, the character of the diabatic states do not depend on ${\bf R}$, such that $\langle \varphi|{\boldsymbol\nabla}_{\bf R}|\phi\rangle=0$. Using $\hat{\mathcal P}=\sum_{\varphi}|\varphi\rangle\langle \varphi|$, the truncated electronic Hamiltonian is $\hat{\mathcal H}_\mathrm{el}=\hat{\mathcal P}\hat{H}_\mathrm{el}\hat{\mathcal P}=\sum_{\varphi} {\mathcal V}_{\varphi \varphi}({\bf R}) |\varphi\rangle\langle \varphi|+\sum_{\varphi\neq\phi} {\mathcal V}_{\varphi \phi}({\bf R}) |\varphi\rangle\langle \phi|$, 
where ${\mathcal V}_{\varphi \phi}({\bf R})=\langle\varphi|\hat{H}_\mathrm{el}|\phi\rangle$ is the diabatic matrix element of $\hat{H}_\mathrm{el}$ (not to be confused by the previously introduced $\hat{\mathcal V} (\hat{\bf x},\hat{\bf p})=\hat{\mathcal P}\hat{V}(\hat{\bf x})\hat{\mathcal P}$). 

The {\it central} result of this letter is reached by splitting the matter Hamiltonian as $\hat{H}_\mathrm{M}=\hat{\bf T}_{\bf R}+\hat{H}_\mathrm{el}$, then through a similar derivation procedure to obtain the following
\begin{align}\label{hc4}
&\hat{\mathcal{H}}_\mathrm{C}=\hat{\mathcal U}^{\dagger}\hat{\mathcal P}\hat{T}_{R}\hat{\mathcal P}\hat{\mathcal U}+\hat{\mathcal U}^{\dagger}\hat{\mathcal P}\hat{H}_\mathrm{el}(\hat{\bf p}_{r},\hat{\bf x})\hat{\mathcal P}\hat{\mathcal U}+\hat{H}_\mathrm{ph}\\
&=\sum_{j\in R}\frac{1}{2m_j}\hat{\mathcal P}\big(\hat{\bf p}_{j}-{\boldsymbol\nabla}_{j}\tilde{\boldsymbol\mu}\hat{\bf A}+{\bf \tilde P}_{j}\big)^2\hat{\mathcal P}+\hat{\mathcal U}^{\dagger}\hat{\mathcal H}_\mathrm{el}\hat{\mathcal U}+\hat{H}_\mathrm{ph},\nonumber
\end{align}
where the sum over $j$ {\it only} includes nuclei. In the above expression, we did not specify the choice of $\hat{\mathcal P}$, which could be either adiabatic or diabatic. Under the limiting case when ${\bf A}_0=0$ or $\tilde{\boldsymbol\mu}\cdot\hat{\bf A}=0$, both the $-{\boldsymbol\nabla}_{j}\tilde{\boldsymbol\mu}\hat{\bf A}$ and ${\bf \tilde P}_{j}$ terms (see Eqn.~\ref{eqn:boostp}) become 0, and $\hat{\mathcal U}^{\dagger}=\hat{\mathcal U}\to \hat{\mathcal P}\otimes\hat{\mathds{1}}_\mathrm{R}\otimes\hat{\mathds{1}}_\mathrm{ph}$. Thus, under a such limit, $\hat{\mathcal{H}}_\mathrm{C}\to\hat{\mathcal H}_\mathrm{M}+\hat{H}_\mathrm{ph}$ hence the matter and the cavity becomes decoupled. When using adiabatic states for the truncation, $\hat{\mathcal P}\hat{\bf p}^2_{j}\hat{\mathcal P}=\sum_{\alpha,\beta}(\hat{\bf p}_{j}\delta_{\alpha\beta}-i\hbar{\bf d}^{j}_{\alpha\beta})^{2} |\alpha\rangle\langle \beta|$, where ${\bf d}^{j}_{\alpha\beta}\equiv \langle \alpha|{\boldsymbol\nabla}_{j}|\beta\rangle$ is the well known derivative couplings. Besides these adiabatic derivative couplings, the light-matter interaction also induced additional ``derivative"-type couplings, $-{\boldsymbol\nabla}_{j}\tilde{\boldsymbol\mu}\hat{\bf A}$ and ${\bf \tilde P}_{j}$, regardless of the electronic representation used in constructing $\hat{\mathcal P}$. When using the Mulliken-Hush diabatic states~\cite{cave1996,cave1997} which are the eigenstates of the $\tilde{\boldsymbol\mu}\equiv \hat{\mathcal P}\hat{\boldsymbol \mu}\hat{\mathcal P}$ operator, such that $\tilde{\boldsymbol\mu}=\sum_{\phi}\mu_{\phi\phi}|\phi\rangle\langle \phi|$, one can prove that $\tilde{\bf P}_{j}=0$ for all nuclei. This is because that ${\boldsymbol\nabla}_{j}\tilde{\boldsymbol\mu}=\sum_{\phi}{\boldsymbol\nabla}_{j}\mu_{\phi\phi}|\phi\rangle\langle \phi|$, thus both $\tilde{\boldsymbol\mu}\hat{\bf A}$ and $[\tilde{\boldsymbol\mu}\hat{\bf A},\hat{\bf p}_j]$ become purely diagonal matrices, hence all of the higher order commutators in Eqn.~\ref{eqn:boostp} become zero, resulting in $\tilde {\bf P}_{j}=0$ for $j\in {\bf R}$.

$\hat{\mathcal{H}}_\mathrm{C}$ in Eqn.~\ref{hc4} provides a general and convenient expression for the Coulomb gauge molecular cavity QED Hamiltonian under the truncation of electronic states. To construct it, one can follow the general procedure: (i) obtain the matter Hamiltonian in this truncated subspace $\hat{\mathcal{H}}_\mathrm{el}=\hat{\mathcal P}\hat{H}_\mathrm{el}\hat{\mathcal P}$ and the dipole operator in this subspace $\tilde{\boldsymbol\mu}=\hat{\mathcal P}\hat{\boldsymbol\mu}\hat{\mathcal P}$ (which in principle include both the transition and permanent dipoles), (ii) constructing the operator $\hat{\mathcal U}=\exp \big[-\frac{i}{\hbar}\hat{\mathcal P}\hat{\boldsymbol\mu}\hat{\mathcal P}\hat{\bf A}\big]$, and obtain $\hat{\mathcal U}^{\dagger}\hat{\mathcal H}_\mathrm{el}\hat{\mathcal U}$, and (iii) evaluating ${\boldsymbol\nabla}_{j}\tilde{\boldsymbol\mu}$, as well as $\tilde{\bf P}_{j}$, then couple them with $\hat{\bf p}_{R}$ and project whole term with $\hat{\mathcal P}$. This procedure should extend to the semi-classical treatment of light-matter interaction when considering a few level matter interacting with an intense classical laser field~\cite{Li2020}, where the substitution $\hat{\bf A}\to {\bf A}(t)$ is often made. In the Coulomb gauge, one need do this replacement in $\hat{\mathcal H}_\mathrm{C}$ instead in $\hat{\mathcal H}'_\mathrm{C}$.

Next, we use the above general principle to derive analytical results for a model system. Without loosing generality, let us consider a molecular system within the diabatic states $\{|0\rangle,|1\rangle\}$, which represents a broad range of chemical systems~\cite{Nitzan2019,Triana2018,Triana2019}. To simplify our algebra, we will assume there is only one nuclear DOF with the coordinate $\hat{R}$ and momentum $\hat{p}_{R}$. We will further assume that $\hat{\boldsymbol\mu}$ is always aligned along the polarization direction of $\hat{\bf A}$ (which is $\hat{\bf e}$). 
Under the truncated space, $\hat{\mathcal P}=|0\rangle\langle 0|+|1\rangle\langle 1|$, the dipole operator is expressed as $\tilde{\mu}\equiv\hat{\mathcal P}\hat{\mu}\hat{\mathcal P}=\Delta\mu\hat{\sigma}_z +\bar{\mu}\hat{\mathcal P} +\mu_{10}\hat{\sigma}_{x}$, where $\Delta\mu=\frac{1}{2} (\mu_{00}-\mu_{11})$, $\bar{\mu}=\frac{1}{2}(\mu_{00}+\mu_{11})$, and $\mu_{\varphi\phi}(\hat{R})=\langle \varphi|\hat{\mu}|\phi\rangle$. Note that these transition and permanent dipoles are functions of $\hat{R}$. The electronic Hamiltonian in this truncated subspace is $\hat{\mathcal H}_\mathrm{el}=\hat{\mathcal P}\hat{H}_\mathrm{el}\hat{\mathcal P}=\varepsilon (\hat{R})\hat{\sigma}_z + \bar{\mathcal V}(\hat{R})\hat{\mathcal P}+ {\mathcal V}_{10} (\hat{R})\hat{\sigma}_x$,
where $\varepsilon (\hat{R}) =\frac{1}{2}({\mathcal V}_{00}(\hat{R})-{\mathcal V}_{11}(\hat{R}))$, $\bar{\mathcal V} (\hat{ R})=\frac{1}{2}({\mathcal V}_{00}(\hat{R})+{\mathcal V}_{11}(\hat{R}))$, and ${\mathcal V}_{\varphi\phi}(\hat{R})=\langle \varphi|\hat{H}_\mathrm{el}|\phi\rangle$ ({\it i.e.}, they are $\hat{H}_\mathrm{el}$'s matrix elements). Using the above spin representation for $\tilde{\mu}$ and $\hat{H}_\mathrm{el}$, as well as the BCH identity, one can analytically show (see details in SI) that for $\hat{\mathcal H}_\mathrm{C}$ in Eqn.~\ref{hc4}, we have 
\begin{align}
&\hat{\mathcal U}^{\dagger}\hat{\mathcal H}_\mathrm{el}\hat{\mathcal U}=\hat{\mathcal H}_\mathrm{el} +\Big(\varepsilon (\hat{R}) \sin\theta - {\mathcal V}_{10}(\hat{R}) \cos\theta\Bigr)\Bigl( \sin{[\xi \hat{A}]} \hat{\sigma}_y  \nonumber \\
&~+\cos{\theta} \bigl(1 - \cos[\xi \hat{A}]\bigr) \hat{\sigma}_x + \sin{\theta} \bigl(\cos[\xi \hat{A}]-1\bigr) \hat{\sigma}_z \Bigr), \label{pzw-hel}
\end{align}
$\xi= \sqrt{(\mu_{00}-\mu_{11})^2 + 4 \mu_{10}^2}$, $\tan\theta= 2 \mu_{01}/(\mu_{00}-\mu_{11})$, and the residual momentum is ${\tilde P}_{R}=\frac{1}{2} \big(\nabla_{R} \tan\theta\big) \cos^2\theta
\big[\big(1-\cos[\xi \hat{A}]\big)\hat{\sigma}_y +\big((\sin\theta) \hat{\sigma}_z - (\cos\theta)\hat{\sigma}_x \big) \big(\sin[\xi \hat{A}] - \xi \hat{A}  \big)\big]$. Note that for using adiabatic states projection $\hat{\mathcal P}=\sum_{\alpha}|\alpha(R)\rangle \langle \alpha (R)|$, the $\hat{\mathcal U}^{\dagger}\hat{\mathcal H}_\mathrm{el}\hat{\mathcal U}$ expression in Eqn.~\ref{pzw-hel} remains the same form with $\hat{\mathcal V}_{01}=0$, and so does the form of $\hat{\mathcal H}_\mathrm{C}$, except for the detailed expression of $\tilde{P}$. The above result has two interesting limits. The first limit is the Rabi model (two-level atom interacts with a cavity) under the Coulomb gauge. For such an atomic-cavity system, there is no nuclear DOF $\{R,p_{R}\}$, and $\hat{\mathcal{H}}_\mathrm{M}=\hat{\mathcal{H}}_\mathrm{el}=\varepsilon\hat{\sigma}_z + \bar{E}\hat{\mathcal P}$, where $\varepsilon =\frac{1}{2}({E}_{0}-{E}_{1})=\frac{1}{2}\hbar\omega_{01}$, $\bar{E}=\frac{1}{2}({E}_{0}+{E}_{1})$, and $\hat{\mathcal P}\hat{\mu}\hat{\mathcal P}= {\mu}_{10}(|0\rangle \langle 1|+|1\rangle \langle 0|)={\mu}_{10}\hat{\sigma}_x$. Further, by noticing that ${\mathcal V}_{10} \rightarrow 0$, $\xi \rightarrow 2\mu_{10}$, $\tan\theta \rightarrow \infty$, $\theta\to \pi/2$, and $\sin\theta \rightarrow 1$, $\cos\theta \rightarrow0$, Equation \ref{pzw-hel} simplifies to $\hat{\mathcal{H}}_\mathrm{C} = \bar{E}\hat{\mathcal P}+\varepsilon\sin[2 \mu_{10} \hat{A}] \hat{\sigma}_y + \varepsilon \cos[2 \mu_{12} \hat{A}] \hat{\sigma}_{z}+\hbar \omega (\hat{a}^\dagger \hat{a} +\frac{1}{2})$, which is the Rabi model under the Coulomb gauge~\cite{Nori2019natphys} that provides the consistent results of the dipole gauge Rabi model. 
The second limit is when these diabatic states are also Mulliken-Hush diabatic states, which means that $\mu_{10}=0$ and $\tan \theta (R)=0$, hence $\nabla_{R}\tan \theta (R)=0$, and $\tilde{P}_{R}=0$, agreeing with our previous analysis of this residual momentum. 

\begin{figure}
 \centering
  \begin{minipage}[h]{1.0\linewidth}
     \centering
     \includegraphics[width=\linewidth]{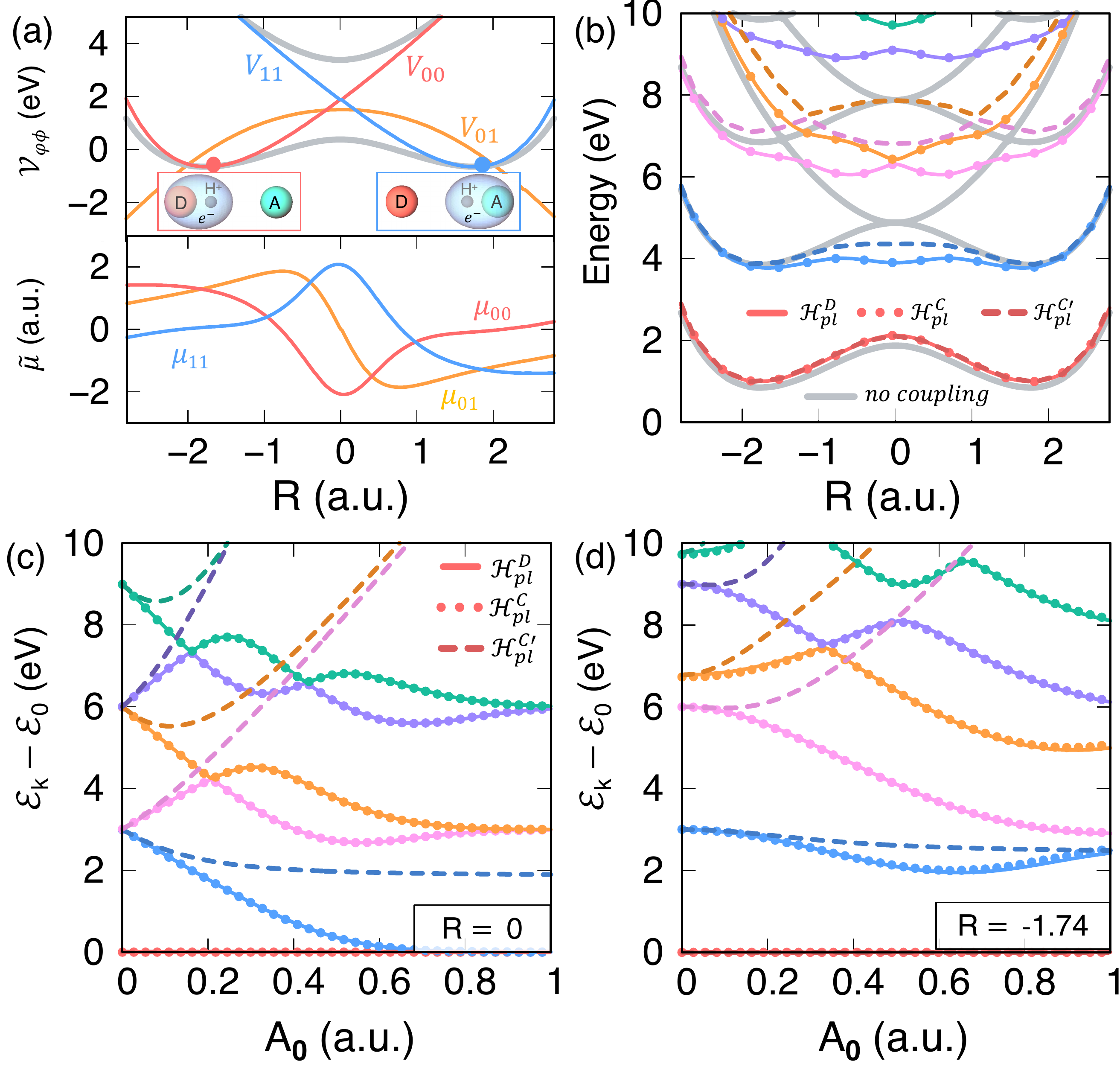}
       \end{minipage}%
   \caption{\small Shin-Metiu model (transferring proton and electron between two fixed ions) coupled to an optical cavity. (a) Diabatic potentials ${\mathcal V}_{\varphi\phi}(R)$ (upper panel) and dipole $\tilde{\mu}$ (lower panel), with the inset describes characters of the diabatic states. (b) Polariton potential energy surface $\mathcal {E}_{k}$ for the molecule-cavity hybrid system at $A_0=0.2$ and $\hbar\omega_\mathrm{c}=3$ eV, from $\hat{\mathcal H}^\mathrm{D}_\mathrm{pl}$ (solid), $\hat{\mathcal H}^\mathrm{C}_\mathrm{pl}$ (dotted), and $\hat{\mathcal H}^\mathrm{C'}_\mathrm{pl}$ (dashed). The polariton eigen-energies as a function of $A_0$ are depicted at (c) R=0 and (d) R=-1.74 a.u.}
\label{Model}
\end{figure}
Fig.~\ref{Model} demonstrates the validity of our theory with a numerical example of a molecule couple to the cavity~\cite{Flick2017PNAS}. Here, we use the Shin-Metiu model~\cite{Metiu:1995} to represent the molecular system, which contains two fixed ions, one moving electron and proton (whose position is $R$), all interacting with each other through modified Coulombic potentials. The details of this model, as well as the procedure to obtain the strict diabatic states (which is not the MH diabatic representation) are provided in SI. Fig.~\ref{Model}a presents the diabatic potential and the matrix elements of $\tilde{\mu}$. Here, we focus on comparing the polaritonic potential energy surface $\mathcal {E}_{k}(R)$, which is defined as $\hat{H}^\mathrm{D}_\mathrm{pl}|\Phi_{k}(R)\rangle=\mathcal {E}_{k}(R)|\Phi_{k}(R)\rangle$ where $\hat{H}^\mathrm{D}_\mathrm{pl}=\hat{H}_\mathrm{D}-\hat{T}_{R}$ represents the polariton Hamiltonian under the dipole gauge. In the truncated electronic subspace, $\hat{\mathcal H}^\mathrm{D}_\mathrm{pl}=\hat{\mathcal H}_\mathrm{el}+\hat{\mathcal U}\hat{H}_\mathrm{ph}\hat{\mathcal U}^{\dagger}$, $\hat{\mathcal H}^\mathrm{C}_\mathrm{pl}=\hat{\mathcal U}^{\dagger} \hat{\mathcal H}^\mathrm{D}_\mathrm{pl}\hat{\mathcal U}=\hat{\mathcal U}^{\dagger}\hat{\mathcal H}_\mathrm{el}\hat{\mathcal U}+\hat{H}_\mathrm{ph}$, and $\hat{\mathcal H}^\mathrm{C'}_\mathrm{pl}=\hat{\mathcal P}\hat{U}^{\dagger} \hat{H}^\mathrm{D}_\mathrm{pl}\hat{U}\hat{\mathcal P}$. Note that the analytical results of $\hat{\mathcal U}^{\dagger}\hat{\mathcal H}_\mathrm{el}\hat{\mathcal U}$ is expressed in Eqn.~\ref{pzw-hel}, whereas the details of other expressions are provided in SI. The matrix elements of these Hamiltonians are evaluated with the two electronic diabatic states and a large number of Fock states, and diagonalizing this matrix gives $\mathcal {E}_{k}(R)$. Fig.~\ref{Model}b presents $\mathcal {E}_{k}(R)$ with $A_0=0.2$ a.u. and $\hbar\omega_\mathrm{c}=3$ eV (such that the light and matter excitations are in resonance at $R=0$). While the $\hat{\mathcal H}^\mathrm{D}_\mathrm{pl}$ (solid) and $\hat{\mathcal H}^\mathrm{C}_\mathrm{pl}$ (dotted) give identical results throughout all range of $R$, $\hat{\mathcal H}^\mathrm{C'}_\mathrm{pl}$ (dashed) gives inconsistent results and breakdown gauge invariance. Fig.~\ref{Model}c-d presents $\mathcal {E}_{k}-\mathcal {E}_{0}$ at $\hbar\omega_\mathrm{c}=3$ eV as a function of the field strength $A_0$, at $R=0$ (resonance condition) and $R=-1.74$ a.u. (detuned), respectively. Again, the results from the Coulomb gauge and dipole gauge agree with each other exactly through out the entire range of the field strength, whereas simple state truncation on the Coulomb gauge QED Hamiltonian breaks the gauge invariance, especially in the ultra-strong coupling regime~\cite{Nori2019natphys,Kockum2019}.

In conclusion, we lay out the fundamental theoretical framework for the molecular cavity QED by presenting the general procedure to obtain the Coulomb gauge Hamiltonian that provide the consistent results from the dipole gauge Hamiltonian, under the same level of electronic state truncation. Using a finite level of projection on $\hat{H}_\mathrm{D}$ often provides accurate results of the polariton eigenspecturm, whereas  $\hat{H}'_\mathrm{C}$ often introduce the gauge ambiguity (especially in the ultra-strong coupling regime). Instead, $\hat{\mathcal H}_\mathrm{C}$ in Eqn.~\ref{hc4} resolves such gauge ambiguity by proving consistent results as $\hat{\mathcal H}_\mathrm{D}$. Investigations based upon the Coulomb gauge~\cite{Rubio2019TMD} should consider using $\hat{\mathcal H}_\mathrm{C}$. 

\begin{acknowledgments}
\section{Acknowledgments}
This work was supported by the National Science Foundation ``Enabling Quantum Leap in Chemistry" program under the Grant number CHE-1836546. M.A.D.T would like to thank Prof. Jim Zavislan and the Institute of Optics for supporting this research as a part of his senior thesis. A.M. appreciates the support from his Elon Huntington Hooker Fellowship. W.Z. appreciates the support from the China Scholarship Council. P. H. acknowledges the support from his Cottrell Scholar award. We appreciate valuable discussions with Prof. Ahsan Nazir.

M.A.D.T and A. M contributed equally to this work.
\end{acknowledgments}

\bibliography{PZW}

\begin{thebibliography}{36}%
\makeatletter
\providecommand \@ifxundefined [1]{%
 \@ifx{#1\undefined}
}%
\providecommand \@ifnum [1]{%
 \ifnum #1\expandafter \@firstoftwo
 \else \expandafter \@secondoftwo
 \fi
}%
\providecommand \@ifx [1]{%
 \ifx #1\expandafter \@firstoftwo
 \else \expandafter \@secondoftwo
 \fi
}%
\providecommand \natexlab [1]{#1}%
\providecommand \enquote  [1]{``#1''}%
\providecommand \bibnamefont  [1]{#1}%
\providecommand \bibfnamefont [1]{#1}%
\providecommand \citenamefont [1]{#1}%
\providecommand \href@noop [0]{\@secondoftwo}%
\providecommand \href [0]{\begingroup \@sanitize@url \@href}%
\providecommand \@href[1]{\@@startlink{#1}\@@href}%
\providecommand \@@href[1]{\endgroup#1\@@endlink}%
\providecommand \@sanitize@url [0]{\catcode `\\12\catcode `\$12\catcode
  `\&12\catcode `\#12\catcode `\^12\catcode `\_12\catcode `\%12\relax}%
\providecommand \@@startlink[1]{}%
\providecommand \@@endlink[0]{}%
\providecommand \url  [0]{\begingroup\@sanitize@url \@url }%
\providecommand \@url [1]{\endgroup\@href {#1}{\urlprefix }}%
\providecommand \urlprefix  [0]{URL }%
\providecommand \Eprint [0]{\href }%
\providecommand \doibase [0]{https://doi.org/}%
\providecommand \selectlanguage [0]{\@gobble}%
\providecommand \bibinfo  [0]{\@secondoftwo}%
\providecommand \bibfield  [0]{\@secondoftwo}%
\providecommand \translation [1]{[#1]}%
\providecommand \BibitemOpen [0]{}%
\providecommand \bibitemStop [0]{}%
\providecommand \bibitemNoStop [0]{.\EOS\space}%
\providecommand \EOS [0]{\spacefactor3000\relax}%
\providecommand \BibitemShut  [1]{\csname bibitem#1\endcsname}%
\let\auto@bib@innerbib\@empty
\bibitem [{\citenamefont {Flick}\ \emph {et~al.}(2017)\citenamefont {Flick},
  \citenamefont {Ruggenthaler}, \citenamefont {Appel},\ and\ \citenamefont
  {Rubio}}]{Flick2017PNAS}%
  \BibitemOpen
  \bibfield  {author} {\bibinfo {author} {\bibfnamefont {J.}~\bibnamefont
  {Flick}}, \bibinfo {author} {\bibfnamefont {M.}~\bibnamefont {Ruggenthaler}},
  \bibinfo {author} {\bibfnamefont {H.}~\bibnamefont {Appel}},\ and\ \bibinfo
  {author} {\bibfnamefont {A.}~\bibnamefont {Rubio}},\ }\bibfield  {title}
  {\bibinfo {title} {Atoms and molecules in cavities, from weak to strong
  coupling in quantum-electrodynamics (qed) chemistry},\ }\href@noop {}
  {\bibfield  {journal} {\bibinfo  {journal} {Proc. Natl. Acad. Sci. U. S. A.}\
  }\textbf {\bibinfo {volume} {114}},\ \bibinfo {pages} {3026} (\bibinfo {year}
  {2017})}\BibitemShut {NoStop}%
\bibitem [{\citenamefont {Ebbesen}(2016)}]{Ebbesen16}%
  \BibitemOpen
  \bibfield  {author} {\bibinfo {author} {\bibfnamefont {T.~W.}\ \bibnamefont
  {Ebbesen}},\ }\bibfield  {title} {\bibinfo {title} {Hybrid light-matter
  states in a molecular and material science perspective},\ }\href@noop {}
  {\bibfield  {journal} {\bibinfo  {journal} {Acc. Chem. Res.}\ }\textbf
  {\bibinfo {volume} {49}},\ \bibinfo {pages} {2403} (\bibinfo {year}
  {2016})}\BibitemShut {NoStop}%
\bibitem [{\citenamefont {Feist}\ \emph {et~al.}(2018)\citenamefont {Feist},
  \citenamefont {Galego},\ and\ \citenamefont {Garcia-Vidal}}]{Feist2018}%
  \BibitemOpen
  \bibfield  {author} {\bibinfo {author} {\bibfnamefont {J.}~\bibnamefont
  {Feist}}, \bibinfo {author} {\bibfnamefont {J.}~\bibnamefont {Galego}},\ and\
  \bibinfo {author} {\bibfnamefont {F.~J.}\ \bibnamefont {Garcia-Vidal}},\
  }\bibfield  {title} {\bibinfo {title} {Polaritonic chemistry with organic
  molecules},\ }\href@noop {} {\bibfield  {journal} {\bibinfo  {journal} {ACS
  Photonics}\ }\textbf {\bibinfo {volume} {5}},\ \bibinfo {pages} {205}
  (\bibinfo {year} {2018})}\BibitemShut {NoStop}%
\bibitem [{\citenamefont {Ribeiro}\ \emph {et~al.}(2018)\citenamefont
  {Ribeiro}, \citenamefont {Martínez-Martínez}, \citenamefont {Du},
  \citenamefont {Campos-Gonzalez-Angulo},\ and\ \citenamefont
  {Yuen-Zhou}}]{Ribeiro2018}%
  \BibitemOpen
  \bibfield  {author} {\bibinfo {author} {\bibfnamefont {R.~F.}\ \bibnamefont
  {Ribeiro}}, \bibinfo {author} {\bibfnamefont {L.~A.}\ \bibnamefont
  {Martínez-Martínez}}, \bibinfo {author} {\bibfnamefont {M.}~\bibnamefont
  {Du}}, \bibinfo {author} {\bibfnamefont {J.}~\bibnamefont
  {Campos-Gonzalez-Angulo}},\ and\ \bibinfo {author} {\bibfnamefont
  {J.}~\bibnamefont {Yuen-Zhou}},\ }\bibfield  {title} {\bibinfo {title}
  {Polariton chemistry: controlling molecular dynamics with optical cavities},\
  }\href@noop {} {\bibfield  {journal} {\bibinfo  {journal} {Chem. Sci.}\
  }\textbf {\bibinfo {volume} {9}},\ \bibinfo {pages} {6325} (\bibinfo {year}
  {2018})}\BibitemShut {NoStop}%
\bibitem [{\citenamefont {Hutchison}\ \emph {et~al.}(2012)\citenamefont
  {Hutchison}, \citenamefont {Schwartz}, \citenamefont {Genet}, \citenamefont
  {Devaux},\ and\ \citenamefont {Ebbesen}}]{Hutchison12}%
  \BibitemOpen
  \bibfield  {author} {\bibinfo {author} {\bibfnamefont {J.~A.}\ \bibnamefont
  {Hutchison}}, \bibinfo {author} {\bibfnamefont {T.}~\bibnamefont {Schwartz}},
  \bibinfo {author} {\bibfnamefont {C.}~\bibnamefont {Genet}}, \bibinfo
  {author} {\bibfnamefont {E.}~\bibnamefont {Devaux}},\ and\ \bibinfo {author}
  {\bibfnamefont {T.~W.}\ \bibnamefont {Ebbesen}},\ }\bibfield  {title}
  {\bibinfo {title} {Modifying chemical landscapes by coupling to vacuum
  fields},\ }\href@noop {} {\bibfield  {journal} {\bibinfo  {journal} {Angew.
  Chem. Int. Ed.}\ }\textbf {\bibinfo {volume} {51}},\ \bibinfo {pages} {1592}
  (\bibinfo {year} {2012})}\BibitemShut {NoStop}%
\bibitem [{\citenamefont {Kowalewski}\ \emph {et~al.}(2016)\citenamefont
  {Kowalewski}, \citenamefont {Bennett},\ and\ \citenamefont
  {Mukamel}}]{KowalewskiJCP2016}%
  \BibitemOpen
  \bibfield  {author} {\bibinfo {author} {\bibfnamefont {M.}~\bibnamefont
  {Kowalewski}}, \bibinfo {author} {\bibfnamefont {K.}~\bibnamefont
  {Bennett}},\ and\ \bibinfo {author} {\bibfnamefont {S.}~\bibnamefont
  {Mukamel}},\ }\bibfield  {title} {\bibinfo {title} {Non-adiabatic dynamics of
  molecules in optical cavities},\ }\href@noop {} {\bibfield  {journal}
  {\bibinfo  {journal} {J. Chem. Phys.}\ }\textbf {\bibinfo {volume} {144}},\
  \bibinfo {pages} {054309} (\bibinfo {year} {2016})}\BibitemShut {NoStop}%
\bibitem [{\citenamefont {Galego}\ \emph {et~al.}(2016)\citenamefont {Galego},
  \citenamefont {Garcia-Vidal},\ and\ \citenamefont {Feist}}]{Galego2016}%
  \BibitemOpen
  \bibfield  {author} {\bibinfo {author} {\bibfnamefont {J.}~\bibnamefont
  {Galego}}, \bibinfo {author} {\bibfnamefont {F.~J.}\ \bibnamefont
  {Garcia-Vidal}},\ and\ \bibinfo {author} {\bibfnamefont {J.}~\bibnamefont
  {Feist}},\ }\bibfield  {title} {\bibinfo {title} {Suppressing photochemical
  reactions with quantized light fields},\ }\href@noop {} {\bibfield  {journal}
  {\bibinfo  {journal} {Nat. Commun.}\ }\textbf {\bibinfo {volume} {7}},\
  \bibinfo {pages} {13841 EP} (\bibinfo {year} {2016})}\BibitemShut {NoStop}%
\bibitem [{\citenamefont {Mandal}\ and\ \citenamefont
  {Huo}(2019)}]{Mandal2019JPCL}%
  \BibitemOpen
  \bibfield  {author} {\bibinfo {author} {\bibfnamefont {A.}~\bibnamefont
  {Mandal}}\ and\ \bibinfo {author} {\bibfnamefont {P.}~\bibnamefont {Huo}},\
  }\bibfield  {title} {\bibinfo {title} {Investigating new reactivities enabled
  by polariton photochemistry},\ }\href@noop {} {\bibfield  {journal} {\bibinfo
   {journal} {J. Phys. Chem. Lett.}\ }\textbf {\bibinfo {volume} {10}},\
  \bibinfo {pages} {5519} (\bibinfo {year} {2019})}\BibitemShut {NoStop}%
\bibitem [{\citenamefont {Thomas}\ \emph {et~al.}(2019)\citenamefont {Thomas},
  \citenamefont {Lethuillier-Karl}, \citenamefont {Nagarajan}, \citenamefont
  {Vergauwe}, \citenamefont {George}, \citenamefont {Chervy}, \citenamefont
  {Shalabney}, \citenamefont {Devaux}, \citenamefont {Genet}, \citenamefont
  {Moran},\ and\ \citenamefont {Ebbesen}}]{Thomas2019}%
  \BibitemOpen
  \bibfield  {author} {\bibinfo {author} {\bibfnamefont {A.}~\bibnamefont
  {Thomas}}, \bibinfo {author} {\bibfnamefont {L.}~\bibnamefont
  {Lethuillier-Karl}}, \bibinfo {author} {\bibfnamefont {K.}~\bibnamefont
  {Nagarajan}}, \bibinfo {author} {\bibfnamefont {R.~M.~A.}\ \bibnamefont
  {Vergauwe}}, \bibinfo {author} {\bibfnamefont {J.}~\bibnamefont {George}},
  \bibinfo {author} {\bibfnamefont {T.}~\bibnamefont {Chervy}}, \bibinfo
  {author} {\bibfnamefont {A.}~\bibnamefont {Shalabney}}, \bibinfo {author}
  {\bibfnamefont {E.}~\bibnamefont {Devaux}}, \bibinfo {author} {\bibfnamefont
  {C.}~\bibnamefont {Genet}}, \bibinfo {author} {\bibfnamefont
  {J.}~\bibnamefont {Moran}},\ and\ \bibinfo {author} {\bibfnamefont {T.~W.}\
  \bibnamefont {Ebbesen}},\ }\bibfield  {title} {\bibinfo {title} {Tilting a
  ground-state reactivity landscape by vibrational strong coupling},\
  }\href@noop {} {\bibfield  {journal} {\bibinfo  {journal} {Science}\ }\textbf
  {\bibinfo {volume} {363}},\ \bibinfo {pages} {615} (\bibinfo {year}
  {2019})}\BibitemShut {NoStop}%
\bibitem [{\citenamefont {Cohen-Tannoudji}\ \emph {et~al.}(1989)\citenamefont
  {Cohen-Tannoudji}, \citenamefont {Dupont-Roc},\ and\ \citenamefont
  {Grynberg}}]{Cohen-Tannoudji}%
  \BibitemOpen
  \bibfield  {author} {\bibinfo {author} {\bibfnamefont {C.}~\bibnamefont
  {Cohen-Tannoudji}}, \bibinfo {author} {\bibfnamefont {J.}~\bibnamefont
  {Dupont-Roc}},\ and\ \bibinfo {author} {\bibfnamefont {G.}~\bibnamefont
  {Grynberg}},\ }\bibfield  {title} {\bibinfo {title} {Photons and atoms:
  Introduction to quantum electrodynamics},\ }\href@noop {} {\bibfield
  {journal} {\bibinfo  {journal} {John Wiley \& Sons, Inc.}\ } (\bibinfo {year}
  {1989})}\BibitemShut {NoStop}%
\bibitem [{\citenamefont {Grynberg}\ \emph {et~al.}(2010)\citenamefont
  {Grynberg}, \citenamefont {Aspect},\ and\ \citenamefont {Fabre}}]{Aspect}%
  \BibitemOpen
  \bibfield  {author} {\bibinfo {author} {\bibfnamefont {G.}~\bibnamefont
  {Grynberg}}, \bibinfo {author} {\bibfnamefont {A.}~\bibnamefont {Aspect}},\
  and\ \bibinfo {author} {\bibfnamefont {C.}~\bibnamefont {Fabre}},\ }\bibfield
   {title} {\bibinfo {title} {Introduction to quantum optics: From the
  semi-classical approach to quantized light},\ }\href@noop {} {\bibfield
  {journal} {\bibinfo  {journal} {Cambridge University Press}\ } (\bibinfo
  {year} {2010})}\BibitemShut {NoStop}%
\bibitem [{\citenamefont {Lamb}(1952)}]{Lamb1952}%
  \BibitemOpen
  \bibfield  {author} {\bibinfo {author} {\bibfnamefont {W.~E.}\ \bibnamefont
  {Lamb}},\ }\bibfield  {title} {\bibinfo {title} {Fine structure of the
  hydrogen atom. iii.},\ }\href@noop {} {\bibfield  {journal} {\bibinfo
  {journal} {Phys. Rev.}\ }\textbf {\bibinfo {volume} {85}},\ \bibinfo {pages}
  {259} (\bibinfo {year} {1952})}\BibitemShut {NoStop}%
\bibitem [{\citenamefont {Aharonov}\ and\ \citenamefont {Au}(1981)}]{Aharonov}%
  \BibitemOpen
  \bibfield  {author} {\bibinfo {author} {\bibfnamefont {Y.}~\bibnamefont
  {Aharonov}}\ and\ \bibinfo {author} {\bibfnamefont {C.}~\bibnamefont {Au}},\
  }\bibfield  {title} {\bibinfo {title} {The question of gauge dependence of
  transition probabilities in quantum mechanics: Facts, myths and
  misunderstandings},\ }\href@noop {} {\bibfield  {journal} {\bibinfo
  {journal} {Phys. Lett. A.}\ }\textbf {\bibinfo {volume} {86}},\ \bibinfo
  {pages} {269} (\bibinfo {year} {1981})}\BibitemShut {NoStop}%
\bibitem [{\citenamefont {Lamb}\ \emph {et~al.}(1987)\citenamefont {Lamb},
  \citenamefont {Schlicher},\ and\ \citenamefont {Scully}}]{Lamb1987}%
  \BibitemOpen
  \bibfield  {author} {\bibinfo {author} {\bibfnamefont {W.~E.}\ \bibnamefont
  {Lamb}}, \bibinfo {author} {\bibfnamefont {R.~R.}\ \bibnamefont
  {Schlicher}},\ and\ \bibinfo {author} {\bibfnamefont {M.~O.}\ \bibnamefont
  {Scully}},\ }\bibfield  {title} {\bibinfo {title} {Matter-field interaction
  in atomic physics and quantum optics},\ }\href@noop {} {\bibfield  {journal}
  {\bibinfo  {journal} {Phys. Rev. A}\ }\textbf {\bibinfo {volume} {36}},\
  \bibinfo {pages} {2763} (\bibinfo {year} {1987})}\BibitemShut {NoStop}%
\bibitem [{\citenamefont {Starace}(1971)}]{Starace}%
  \BibitemOpen
  \bibfield  {author} {\bibinfo {author} {\bibfnamefont {A.~F.}\ \bibnamefont
  {Starace}},\ }\bibfield  {title} {\bibinfo {title} {Length and velocity
  formulas in approximate oscillator-strength calculations},\ }\href@noop {}
  {\bibfield  {journal} {\bibinfo  {journal} {Phys. Rev. A}\ }\textbf {\bibinfo
  {volume} {3}},\ \bibinfo {pages} {1242–1245} (\bibinfo {year}
  {1971})}\BibitemShut {NoStop}%
\bibitem [{\citenamefont {Rzazewski}\ and\ \citenamefont {Boyd}(2004)}]{Boyd}%
  \BibitemOpen
  \bibfield  {author} {\bibinfo {author} {\bibfnamefont {K.}~\bibnamefont
  {Rzazewski}}\ and\ \bibinfo {author} {\bibfnamefont {R.}~\bibnamefont
  {Boyd}},\ }\bibfield  {title} {\bibinfo {title} {Equivalence of interaction
  hamiltonians in the electric dipole approximation},\ }\href@noop {}
  {\bibfield  {journal} {\bibinfo  {journal} {Journal of Modern Optics}\
  }\textbf {\bibinfo {volume} {51}},\ \bibinfo {pages} {1137} (\bibinfo {year}
  {2004})}\BibitemShut {NoStop}%
\bibitem [{\citenamefont {Bernardis}\ \emph {et~al.}(2018)\citenamefont
  {Bernardis}, \citenamefont {Pilar}, \citenamefont {Jaako}, \citenamefont
  {Liberato},\ and\ \citenamefont {Rabl}}]{Rabl2018PRA2}%
  \BibitemOpen
  \bibfield  {author} {\bibinfo {author} {\bibfnamefont {D.~D.}\ \bibnamefont
  {Bernardis}}, \bibinfo {author} {\bibfnamefont {P.}~\bibnamefont {Pilar}},
  \bibinfo {author} {\bibfnamefont {T.}~\bibnamefont {Jaako}}, \bibinfo
  {author} {\bibfnamefont {S.~D.}\ \bibnamefont {Liberato}},\ and\ \bibinfo
  {author} {\bibfnamefont {P.}~\bibnamefont {Rabl}},\ }\bibfield  {title}
  {\bibinfo {title} {Breakdown of gauge invariance in ultrastrong-coupling
  cavity qed},\ }\href@noop {} {\bibfield  {journal} {\bibinfo  {journal}
  {Phys. Rev. A}\ }\textbf {\bibinfo {volume} {98}},\ \bibinfo {pages} {053819}
  (\bibinfo {year} {2018})}\BibitemShut {NoStop}%
\bibitem [{\citenamefont {Stefano}\ \emph {et~al.}(2019)\citenamefont
  {Stefano}, \citenamefont {Settineri}, \citenamefont {Macri}, \citenamefont
  {Garziano}, \citenamefont {Stassi}, \citenamefont {Savasta},\ and\
  \citenamefont {Nori}}]{Nori2019natphys}%
  \BibitemOpen
  \bibfield  {author} {\bibinfo {author} {\bibfnamefont {O.~D.}\ \bibnamefont
  {Stefano}}, \bibinfo {author} {\bibfnamefont {A.}~\bibnamefont {Settineri}},
  \bibinfo {author} {\bibfnamefont {V.}~\bibnamefont {Macri}}, \bibinfo
  {author} {\bibfnamefont {L.}~\bibnamefont {Garziano}}, \bibinfo {author}
  {\bibfnamefont {R.}~\bibnamefont {Stassi}}, \bibinfo {author} {\bibfnamefont
  {S.}~\bibnamefont {Savasta}},\ and\ \bibinfo {author} {\bibfnamefont
  {F.}~\bibnamefont {Nori}},\ }\bibfield  {title} {\bibinfo {title} {Resolution
  of gauge ambiguities in ultrastrong-coupling cavity quantum
  electrodynamics},\ }\href@noop {} {\bibfield  {journal} {\bibinfo  {journal}
  {Nature Phys.}\ }\textbf {\bibinfo {volume} {15}},\ \bibinfo {pages} {803}
  (\bibinfo {year} {2019})}\BibitemShut {NoStop}%
\bibitem [{\citenamefont {Stokes}\ and\ \citenamefont {Nazir}(2019)}]{Nazir}%
  \BibitemOpen
  \bibfield  {author} {\bibinfo {author} {\bibfnamefont {A.}~\bibnamefont
  {Stokes}}\ and\ \bibinfo {author} {\bibfnamefont {A.}~\bibnamefont {Nazir}},\
  }\bibfield  {title} {\bibinfo {title} {Gauge ambiguities imply
  jaynes–cummings physics remains valid in ultrastrong coupling qed},\
  }\href@noop {} {\bibfield  {journal} {\bibinfo  {journal} {Nat. Commun.}\
  }\textbf {\bibinfo {volume} {10}},\ \bibinfo {pages} {499} (\bibinfo {year}
  {2019})}\BibitemShut {NoStop}%
\bibitem [{\citenamefont {Power}\ and\ \citenamefont {Zienau}(1959)}]{PZW}%
  \BibitemOpen
  \bibfield  {author} {\bibinfo {author} {\bibfnamefont {E.~A.}\ \bibnamefont
  {Power}}\ and\ \bibinfo {author} {\bibfnamefont {S.}~\bibnamefont {Zienau}},\
  }\bibfield  {title} {\bibinfo {title} {Coulomb gauge in non-relativistic
  quantum electro-dynamics and the shape of spectral lines},\ }\href@noop {}
  {\bibfield  {journal} {\bibinfo  {journal} {Philosophical Transactions of the
  Royal Society of London A, Mathematical and Physical Sciences}\ }\textbf
  {\bibinfo {volume} {251}},\ \bibinfo {pages} {427} (\bibinfo {year}
  {1959})}\BibitemShut {NoStop}%
\bibitem [{\citenamefont {G$\ddot{\mathrm o}$ppert-Mayer}(2009)}]{GM-gauge}%
  \BibitemOpen
  \bibfield  {author} {\bibinfo {author} {\bibfnamefont {M.}~\bibnamefont
  {G$\ddot{\mathrm o}$ppert-Mayer}},\ }\bibfield  {title} {\bibinfo {title}
  {Elementary processes with two quantum transitions},\ }\href@noop {}
  {\bibfield  {journal} {\bibinfo  {journal} {Ann. Phys. (Berlin)}\ }\textbf
  {\bibinfo {volume} {18}},\ \bibinfo {pages} {466} (\bibinfo {year}
  {2009})}\BibitemShut {NoStop}%
\bibitem [{\citenamefont {Frisk~Kockum}\ \emph {et~al.}(2019)\citenamefont
  {Frisk~Kockum}, \citenamefont {Miranowicz}, \citenamefont {De~Liberato},
  \citenamefont {Savasta},\ and\ \citenamefont {Nori}}]{Kockum2019}%
  \BibitemOpen
  \bibfield  {author} {\bibinfo {author} {\bibfnamefont {A.}~\bibnamefont
  {Frisk~Kockum}}, \bibinfo {author} {\bibfnamefont {A.}~\bibnamefont
  {Miranowicz}}, \bibinfo {author} {\bibfnamefont {S.}~\bibnamefont
  {De~Liberato}}, \bibinfo {author} {\bibfnamefont {S.}~\bibnamefont
  {Savasta}},\ and\ \bibinfo {author} {\bibfnamefont {F.}~\bibnamefont
  {Nori}},\ }\bibfield  {title} {\bibinfo {title} {Ultrastrong coupling between
  light and matter},\ }\href@noop {} {\bibfield  {journal} {\bibinfo  {journal}
  {Nat. Rev. Phys.}\ }\textbf {\bibinfo {volume} {1}},\ \bibinfo {pages} {19}
  (\bibinfo {year} {2019})}\BibitemShut {NoStop}%
\bibitem [{\citenamefont {Li}\ \emph {et~al.}(2020)\citenamefont {Li},
  \citenamefont {Golez}, \citenamefont {Mazza}, \citenamefont {Millis},
  \citenamefont {Georges},\ and\ \citenamefont {Eckstein}}]{Li2020}%
  \BibitemOpen
  \bibfield  {author} {\bibinfo {author} {\bibfnamefont {J.}~\bibnamefont
  {Li}}, \bibinfo {author} {\bibfnamefont {D.}~\bibnamefont {Golez}}, \bibinfo
  {author} {\bibfnamefont {G.}~\bibnamefont {Mazza}}, \bibinfo {author}
  {\bibfnamefont {A.~J.}\ \bibnamefont {Millis}}, \bibinfo {author}
  {\bibfnamefont {A.}~\bibnamefont {Georges}},\ and\ \bibinfo {author}
  {\bibfnamefont {M.}~\bibnamefont {Eckstein}},\ }\bibfield  {title} {\bibinfo
  {title} {Electromagnetic coupling in tight-binding models for strongly
  correlated light and matter},\ }\href@noop {} {\bibfield  {journal} {\bibinfo
   {journal} {Phys. Rev. B.}\ }\textbf {\bibinfo {volume} {101}},\ \bibinfo
  {pages} {205140} (\bibinfo {year} {2020})}\BibitemShut {NoStop}%
\bibitem [{\citenamefont {Garziano}\ \emph {et~al.}(2020)\citenamefont
  {Garziano}, \citenamefont {Settineri}, \citenamefont {Stefano}, \citenamefont
  {Savasta},\ and\ \citenamefont {Nori1}}]{Nori2020}%
  \BibitemOpen
  \bibfield  {author} {\bibinfo {author} {\bibfnamefont {L.}~\bibnamefont
  {Garziano}}, \bibinfo {author} {\bibfnamefont {A.}~\bibnamefont {Settineri}},
  \bibinfo {author} {\bibfnamefont {O.~D.}\ \bibnamefont {Stefano}}, \bibinfo
  {author} {\bibfnamefont {S.}~\bibnamefont {Savasta}},\ and\ \bibinfo {author}
  {\bibfnamefont {F.}~\bibnamefont {Nori1}},\ }\bibfield  {title} {\bibinfo
  {title} {Gauge invariance of the dicke and hopfeld models},\ }\href@noop {}
  {\bibfield  {journal} {\bibinfo  {journal} {arXiv}\ ,\ \bibinfo {pages}
  {2002.04241v1}} (\bibinfo {year} {2020})}\BibitemShut {NoStop}%
\bibitem [{\citenamefont {Mead}\ and\ \citenamefont
  {Truhlar}(1982)}]{Mead:1982}%
  \BibitemOpen
  \bibfield  {author} {\bibinfo {author} {\bibfnamefont {C.~A.}\ \bibnamefont
  {Mead}}\ and\ \bibinfo {author} {\bibfnamefont {D.~G.}\ \bibnamefont
  {Truhlar}},\ }\bibfield  {title} {\bibinfo {title} {Conditions for the
  definition of a strictly diabatic electronic basis for molecular systems},\
  }\href@noop {} {\bibfield  {journal} {\bibinfo  {journal} {J. Chem. Phys.}\
  }\textbf {\bibinfo {volume} {77}},\ \bibinfo {pages} {6090} (\bibinfo {year}
  {1982})}\BibitemShut {NoStop}%
\bibitem [{\citenamefont {T.~Pacher}\ and\ \citenamefont
  {Koppel}(1989)}]{Cederbaum:1989}%
  \BibitemOpen
  \bibfield  {author} {\bibinfo {author} {\bibfnamefont {L.~S.~C.}\
  \bibnamefont {T.~Pacher}, \bibfnamefont {C.~A.~Mead}}\ and\ \bibinfo {author}
  {\bibfnamefont {H.}~\bibnamefont {Koppel}},\ }\bibfield  {title} {\bibinfo
  {title} {Gauge theory and quasidiabatic states in molecular physics},\
  }\href@noop {} {\bibfield  {journal} {\bibinfo  {journal} {J. Chem. Phys.}\
  }\textbf {\bibinfo {volume} {91}},\ \bibinfo {pages} {7057} (\bibinfo {year}
  {1989})}\BibitemShut {NoStop}%
\bibitem [{\citenamefont {Cave}\ and\ \citenamefont {Newton}(1996)}]{cave1996}%
  \BibitemOpen
  \bibfield  {author} {\bibinfo {author} {\bibfnamefont {R.~J.}\ \bibnamefont
  {Cave}}\ and\ \bibinfo {author} {\bibfnamefont {M.~D.}\ \bibnamefont
  {Newton}},\ }\bibfield  {title} {\bibinfo {title} {Generalization of the
  mulliken-hush treatment for the calculation of electron transfer matrix
  elements},\ }\href@noop {} {\bibfield  {journal} {\bibinfo  {journal} {Chem.
  Phys. Lett.}\ }\textbf {\bibinfo {volume} {249}},\ \bibinfo {pages} {15 }
  (\bibinfo {year} {1996})}\BibitemShut {NoStop}%
\bibitem [{\citenamefont {Guo}\ and\ \citenamefont
  {Yarkony}(2016)}]{Yarkony:2016}%
  \BibitemOpen
  \bibfield  {author} {\bibinfo {author} {\bibfnamefont {H.}~\bibnamefont
  {Guo}}\ and\ \bibinfo {author} {\bibfnamefont {D.~R.}\ \bibnamefont
  {Yarkony}},\ }\bibfield  {title} {\bibinfo {title} {Accurate nonadiabatic
  dynamics},\ }\href@noop {} {\bibfield  {journal} {\bibinfo  {journal} {Phys.
  Chem. Chem. Phys.}\ }\textbf {\bibinfo {volume} {18}},\ \bibinfo {pages}
  {26335} (\bibinfo {year} {2016})}\BibitemShut {NoStop}%
\bibitem [{\citenamefont {Subotnik}\ \emph {et~al.}(2015)\citenamefont
  {Subotnik}, \citenamefont {Alguire}, \citenamefont {Ou}, \citenamefont
  {Landry},\ and\ \citenamefont {Fatehi}}]{SubotnikDIA}%
  \BibitemOpen
  \bibfield  {author} {\bibinfo {author} {\bibfnamefont {J.~E.}\ \bibnamefont
  {Subotnik}}, \bibinfo {author} {\bibfnamefont {E.~C.}\ \bibnamefont
  {Alguire}}, \bibinfo {author} {\bibfnamefont {Q.}~\bibnamefont {Ou}},
  \bibinfo {author} {\bibfnamefont {B.~R.}\ \bibnamefont {Landry}},\ and\
  \bibinfo {author} {\bibfnamefont {S.}~\bibnamefont {Fatehi}},\ }\bibfield
  {title} {\bibinfo {title} {The requisite electronic structure theory to
  describe photoexcited nonadiabatic dynamics: Nonadiabatic derivative
  couplings and diabatic electronic couplings},\ }\href@noop {} {\bibfield
  {journal} {\bibinfo  {journal} {Acc. Chem. Res.}\ }\textbf {\bibinfo {volume}
  {48}},\ \bibinfo {pages} {1340} (\bibinfo {year} {2015})}\BibitemShut
  {NoStop}%
\bibitem [{\citenamefont {van Voorhis}\ \emph {et~al.}(2010)\citenamefont {van
  Voorhis}, \citenamefont {Kowalczyk}, \citenamefont {Kaduk}, \citenamefont
  {Wang}, \citenamefont {Cheng},\ and\ \citenamefont {Wu}}]{DiabaticARPC}%
  \BibitemOpen
  \bibfield  {author} {\bibinfo {author} {\bibfnamefont {T.}~\bibnamefont {van
  Voorhis}}, \bibinfo {author} {\bibfnamefont {T.}~\bibnamefont {Kowalczyk}},
  \bibinfo {author} {\bibfnamefont {B.}~\bibnamefont {Kaduk}}, \bibinfo
  {author} {\bibfnamefont {L.-P.}\ \bibnamefont {Wang}}, \bibinfo {author}
  {\bibfnamefont {C.-L.}\ \bibnamefont {Cheng}},\ and\ \bibinfo {author}
  {\bibfnamefont {Q.}~\bibnamefont {Wu}},\ }\bibfield  {title} {\bibinfo
  {title} {The diabatic picture of electron transfer, reaction barriers, and
  molecular dynamics},\ }\href@noop {} {\bibfield  {journal} {\bibinfo
  {journal} {Annu. Rev. Phys. Chem.}\ }\textbf {\bibinfo {volume} {61}},\
  \bibinfo {pages} {149} (\bibinfo {year} {2010})}\BibitemShut {NoStop}%
\bibitem [{\citenamefont {Cave}\ and\ \citenamefont {Newton}(1997)}]{cave1997}%
  \BibitemOpen
  \bibfield  {author} {\bibinfo {author} {\bibfnamefont {R.~J.}\ \bibnamefont
  {Cave}}\ and\ \bibinfo {author} {\bibfnamefont {M.~D.}\ \bibnamefont
  {Newton}},\ }\bibfield  {title} {\bibinfo {title} {Calculation of electronic
  coupling matrix elements for ground and excited state electron transfer
  reactions: Comparison of the generalized mulliken–hush and block
  diagonalization methods},\ }\href@noop {} {\bibfield  {journal} {\bibinfo
  {journal} {J. Chem. Phys.}\ }\textbf {\bibinfo {volume} {106}},\ \bibinfo
  {pages} {9213} (\bibinfo {year} {1997})}\BibitemShut {NoStop}%
\bibitem [{\citenamefont {Semenov}\ and\ \citenamefont
  {Nitzan}(2019)}]{Nitzan2019}%
  \BibitemOpen
  \bibfield  {author} {\bibinfo {author} {\bibfnamefont {A.}~\bibnamefont
  {Semenov}}\ and\ \bibinfo {author} {\bibfnamefont {A.}~\bibnamefont
  {Nitzan}},\ }\bibfield  {title} {\bibinfo {title} {Electron transfer in
  confined electromagnetic fields},\ }\href@noop {} {\bibfield  {journal}
  {\bibinfo  {journal} {J. Chem. Phys.}\ }\textbf {\bibinfo {volume} {150}},\
  \bibinfo {pages} {174122} (\bibinfo {year} {2019})}\BibitemShut {NoStop}%
\bibitem [{\citenamefont {Triana}\ \emph {et~al.}(2018)\citenamefont {Triana},
  \citenamefont {Pel\'{a}ez},\ and\ \citenamefont {Sanz-Vicario}}]{Triana2018}%
  \BibitemOpen
  \bibfield  {author} {\bibinfo {author} {\bibfnamefont {J.~F.}\ \bibnamefont
  {Triana}}, \bibinfo {author} {\bibfnamefont {D.}~\bibnamefont {Pel\'{a}ez}},\
  and\ \bibinfo {author} {\bibfnamefont {J.~L.}\ \bibnamefont {Sanz-Vicario}},\
  }\bibfield  {title} {\bibinfo {title} {Entangled photonic-nuclear molecular
  dynamics of lif in quantum optical cavities},\ }\href@noop {} {\bibfield
  {journal} {\bibinfo  {journal} {J. Phys. Chem. A}\ }\textbf {\bibinfo
  {volume} {122}},\ \bibinfo {pages} {2266} (\bibinfo {year}
  {2018})}\BibitemShut {NoStop}%
\bibitem [{\citenamefont {Triana}\ and\ \citenamefont
  {Sanz-Vicario}(2019)}]{Triana2019}%
  \BibitemOpen
  \bibfield  {author} {\bibinfo {author} {\bibfnamefont {J.~F.}\ \bibnamefont
  {Triana}}\ and\ \bibinfo {author} {\bibfnamefont {J.~L.}\ \bibnamefont
  {Sanz-Vicario}},\ }\bibfield  {title} {\bibinfo {title} {Revealing the
  presence of potential crossings in diatomics induced by quantum cavity
  radiation},\ }\href@noop {} {\bibfield  {journal} {\bibinfo  {journal} {Phys.
  Rev. Lett.}\ }\textbf {\bibinfo {volume} {122}},\ \bibinfo {pages} {063603}
  (\bibinfo {year} {2019})}\BibitemShut {NoStop}%
\bibitem [{\citenamefont {Shin}\ and\ \citenamefont
  {Metiu}(1995)}]{Metiu:1995}%
  \BibitemOpen
  \bibfield  {author} {\bibinfo {author} {\bibfnamefont {S.}~\bibnamefont
  {Shin}}\ and\ \bibinfo {author} {\bibfnamefont {H.}~\bibnamefont {Metiu}},\
  }\bibfield  {title} {\bibinfo {title} {{Nonadiabatic Effects on the Charge
  Transfer Rate Constant: A Numerical Study of a Simple Model System}},\
  }\href@noop {} {\bibfield  {journal} {\bibinfo  {journal} {J. Chem. Phys.}\
  }\textbf {\bibinfo {volume} {102}},\ \bibinfo {pages} {9285} (\bibinfo {year}
  {1995})}\BibitemShut {NoStop}%
\bibitem [{\citenamefont {Latini}\ \emph {et~al.}(2019)\citenamefont {Latini},
  \citenamefont {Ronca}, \citenamefont {Giovannini}, \citenamefont
  {H\"{u}bene},\ and\ \citenamefont {Rubio}}]{Rubio2019TMD}%
  \BibitemOpen
  \bibfield  {author} {\bibinfo {author} {\bibfnamefont {S.}~\bibnamefont
  {Latini}}, \bibinfo {author} {\bibfnamefont {E.}~\bibnamefont {Ronca}},
  \bibinfo {author} {\bibfnamefont {U.~D.}\ \bibnamefont {Giovannini}},
  \bibinfo {author} {\bibfnamefont {H.}~\bibnamefont {H\"{u}bene}},\ and\
  \bibinfo {author} {\bibfnamefont {A.}~\bibnamefont {Rubio}},\ }\bibfield
  {title} {\bibinfo {title} {Cavity control of excitons in two-dimensional
  materials},\ }\href@noop {} {\bibfield  {journal} {\bibinfo  {journal} {Nano
  Lett.}\ }\textbf {\bibinfo {volume} {19}},\ \bibinfo {pages} {3473} (\bibinfo
  {year} {2019})}\BibitemShut {NoStop}%
\end{thebibliography}%

\end{document}